\documentclass{article}
\usepackage{ijcai19}

\usepackage{times}
\usepackage{soul}
\usepackage{url}
\usepackage[hidelinks]{hyperref}
\usepackage[utf8]{inputenc}
\usepackage[small]{caption}
\usepackage{graphicx}
\usepackage{amsmath}
\usepackage{booktabs}
\usepackage[ruled]{algorithm2e}
\usepackage[noend]{algpseudocode}

\usepackage{autobreak}

\usepackage{color,soul}

\usepackage{dsfont}
\usepackage{bm}
\usepackage[inline]{enumitem}
\usepackage{subcaption}
\usepackage{multirow}
\usepackage{amsfonts}

\newtheorem{definition}{Definition}
\usepackage{adjustbox}

\newcommand{\citet}[1]{\citeauthor{#1}~\shortcite{#1}}
\newcommand{\citep}{\cite}

\definecolor{orange}{rgb}{1,0.5,0}

\title{Infer Your Enemies and Know Yourself, Learning in Real-Time Bidding with Partially Observable Opponents}

\author{
Manxing Du$^1$\footnote{Work conducted during a research visit at MediaGamma}\and
Alexander I. Cowen-Rivers$^2$\and
Ying Wen$^{2,3}$\and
Phu Sakulwongtana$^{2,3}$\and
Jun Wang$^{2,3}$\and
Mats Brorsson$^4$\And
Radu State$^1$\\
\affiliations
$^1$ University of Luxembourg\\
$^2$ MediaGamma Ltd\\
$^3$ University College London\\
$^4$ Royal Institute of Technology\\
\emails
manxing.du@uni.lu,
alexander.cowen-rivers@mediagamma.com
}

\begin{document}
\maketitle
\begin{abstract}
Real-time bidding, as one of the most popular mechanisms for selling online ad slots, facilitates advertisers to reach their potential customers. The goal of bidding optimization is to maximize the advertisers' return on investment (ROI) under a certain budget setting. A straightforward solution is to model the bidding function in an explicit form. However, the static functional solutions lack generality in practice and are insensitive to the stochastic behaviour of other bidders in the environment. In this paper, we propose a general multi-agent framework with actor-critic solutions facing against playing imperfect information games. We firstly introduce a novel Deep Attentive Survival Analysis (DASA) model to infer the censored data in the second price auctions which outperforms start-of-the-art survival analysis. Furthermore, our approach introduces the DASA model as the opponent model into the policy learning process for each agent and develop a mean field equilibrium analysis of the second price auctions. The experiments have shown that with the inference of the market, the market converges to the equilibrium much faster while playing against both fixed strategy agents and dynamic learning agents.
\end{abstract}

%

%

\section{Introduction}\label{sec:intro}
Real-time bidding (RTB) is a leading online ad inventory trading mechanism in which each ad display is sold through real-time auctions. It allows the advertisers to target potential users with click or purchase interests on the impression level. In RTB, each bidding agent is involved in a highly dynamic bidding environment with an unknown number of competitors. In real-time bidding, second price auctions \cite{krishna2009auction} are usually held where the winner is the one with the highest bid and pays the second highest price. In theory, to reach the Nash equilibrium of the static second price auction, bidders are encouraged to submit their estimation of impression's true value as the bid price \cite{krishna2009auction}. However, in practice, the market may not always maintain the ideal equilibrium due to various reasons. For instance, bidders are constrained by their budget. To avoid running out of money quickly without observing more valuable impressions, the optimal bid price usually deviates from its true value. In addition, the number of participants in each auction is unknown and from each bidder's perspective, it may compete with different opponents at every step during its lifetime. To obtain the optimal bidding strategy in such a stochastic game with a large number of unknown participants is the major challenge in RTB. 

The dynamics in the RTB market mostly come from the hybrid behavior of each bidder. It is important for an intelligent bidding agent to inference its opponents' strategy while optimizing its own strategy. In \cite{he2016opponent}, a deep-Q-Network (DQN) based multi-task reinforcement learning architecture is proposed to jointly learn a policy for an agent and the behavior of its opponents in a multi-player game. The model requires features from opponents as input. Similar ideas have been studied in the RTB domain \cite{jin2018real}, the authors investigate the optimal bidding strategy in a fully observable multi-agent bidding environment, where each agent knows each other's budget and obtained reward at every step. However, in practice, each bidder is not aware of the configuration of its competitors and the competitor set varies in every auction. 

With the number of opponents increasing, modeling every opponent's action becomes implausible and computationally expensive. In addition, given the nature of the second price auctions, only part of the opponent's actions can be observed by each agent. To analyze such highly dynamic games with incomplete information and large number of participants, the Mean Field Theory \cite{stanley1971phase} has been employed. 
A scalable policy learning solution for multi-player games is proposed in \cite{yang2018mean}. The core idea is to find the optimal actions for one agent in response to the mean action of its neighbors. In this way, instead of modeling actions from all the agents in the environment, the mean action of $N$ neighbors represents the action distribution of all neighboring agents. In addition, in ~\cite{gummadi2012repeated,iyer2011mean}, the authors extensively analyzed and proved the existence of the Mean Field Equilibrium (MFE) in the dynamic bidding environment.

Inspired by the prior studies, we firstly address the partially observable opponent actions by adopting a Deep Attentive Survival model which greatly outperforms the state-of-the-art survival models. Furthermore, our solution integrates the opponent model into the policy learning framework for actor-critic based bidding agents. We take the second highest price as the aggregated action of all the other bidders, which enables each bidder to optimize its policy together with modeling the uncertainty of the market. Our experiments have shown the equilibrium under different budget constraints and faster convergence in the multi-agent environment.
 

\section{Related Work}
Bid optimization is one of the key components in the decision making process in RTB. It aims to optimize the bid price on the impression level which maximizes the potential profits under a certain budget constraint. Many research works have formulated it as a functional optimization problem \cite{perlich2012bid,zhang2014optimal}
However, the functional based methods have strong assumptions of the model form and fail to incorporate the dynamics in the bidding environment and the bidder's budget spending status into the model.

To address the above shortage in the prior studies, much research efforts have been focused on Reinforcement Learning (RL) based RTB~\cite{cai2017real,du2017improving,wu2018budget}. These studies mainly address the bidding optimization problem for a single agent, neglecting the stochastic behaviors of other bidders in the market. ~\citet{jin2018real} and ~\citet{zhao2018deep} extend the single agent learning in RTB to a multi-agent bidding scenario. ~\citet{jin2018real} adopt the Deep Deterministic Policy Gradients (DDPG) algorithm on the advertisers cluster level and demonstrate the profit gain per bidder under the competing or collaborating reward settings. However, one strong assumption in this work is that each agent knows each other's state. In practice, the only information each bidder has about other opponents are the market price in case of winning. In the second price auctions, the winning price of the lost auctions are censored which makes the bidding environment to be partially observable. In this case, the bidder only knows the market price should be higher than its own bid price.

To address the censorship problem, survival model has been widely used in the medical domain to conduct the time-to-death analysis \cite{miller2011survival}. The market price estimation has been commonly addressed by adopting a non-parametric KaplanMeier estimator on an aggregated level ~\cite{wang2016functional}. However, one aggregated distribution for all the bid requests fails to capture the divergence in the data. In ~\cite{ren2019deep}, the authors proposed to adopt the recurrent network to model the sequential pattern in the feature space of the individual user and directly estimate the market price probability. However, the features may not only limited to the sequential dependencies. 

The Transformer~\cite{vaswani2017attention} is the first model relying entirely on self-attention to compute a representations of its input without using convolutions or sequence aligned recurrent neural networks~\cite{graves2013speech}. It has fueled much of the latest advancements, such as pre-trained contextualized word embeddings~\cite{peters2018deep,devlin2018bert,radford2018improving} crucial to the success of sequential tasks in natural language processing. In this work, we adopt the transformer model as a non-linear approximation to the survival function as our opponent model.

In the multi-agent stochastic game, for each agent, it is essential to model its opponent's actions. 
The opponent actions are usually either modeled as i.i.d~\cite{brown1951iterative} or as sequential actions with short-term history~\cite{mealing2013opponent}. 
~\citet{hernandez2017learning} assume the opponent redraw strategies during a two-players repeated stochastic game and the agent updates the belief of the opponent model by its observations.
Similar to the repeated stochastic game setting in ~\cite{hernandez2017learning}, our work focuses on repeated second price auctions of the same ad campaign with unknown opponents. The key difference is that we model RTB auctions as a multi-player stochastic game and the opponents are not restricted to have limited memory bounding.

In the repeated auctions, the existence of Mean Field Equilibrium (MFE) under budget constraints has been theoretically proved in~\cite{iyer2014mean,gummadi2012repeated}. Both studies showed that the value function for an agent to reach the MFE takes the known fixed market price, budget, and the observed utility distribution, for example, the estimated click through rate (CTR) as inputs. In practice, the market price is only partially observable to each bidder. In addition, the conventional RL learning algorithms like DDPG, do not explicitly model the opponent'ss strategy. Therefore, in our work, we firstly extend the MFE setting into the Q function estimation in the DDPG algorithm. The opponent model is integrated into Q values using an indicator function, enforcing gradients to only flow through the actions which will result in a reward over the estimated opponents actions. We demonstrate the performance improvements of the optimal MFE strategy from a single agent's perspective.
\section{Problem Formulation}
In this section, we formulate the sequential second price auctions as a multi-player stochastic game. 
Under the classic RL setting, real time bidding process is usually formalized as a Markov Decision Process (MDP) \cite{cai2017real}, which is defined by four elements $(S, A, T, R)$. A state $s_{i} \in S$ describes the status of the agent at step $i$. The bid price is usually considered as the action $a_{i} \in A$ to take. However, the bid price can be set to any number in the range of $(0, \infty)$. To generalize and limit the range of the action $a_{i}$, like in \cite{jin2018real}, in this paper, $a_{i}$ is normalized ranging from $[0, 1]$. The optimal policy $\pi$ is a mapping from $S$ to $A$ which optimizes the reward function $R$ of the agent, in which $r_{i} \in S \times A \mapsto R$. The transition function $T$ defines the probability distributions over the state space: $T: S \times A \mapsto \Omega(S)$.


Different from the conventional RL, where a single agent learns to react to the environment, a stochastic game describes the strategic reaction of all the agents in the environment. In such games, all the agents take actions at the same time, and their actions influence the complex change of the environment. As defined in a stochastic game, at each time step, all the agents choose their part of the joint action $\boldsymbol{a} \in A(\boldsymbol{s})$, where $\boldsymbol{s}$ is the overall game state. In a two players game, for example, the joint action is defined as $\boldsymbol{a} = (a_{i}, a_{-i})$ where $a_{i}$ is the action of player $i$ and $a_{-i}$ is the action of its opponent. The immediate reward $r_{i}$ is represented as $r_{i}(\boldsymbol{s}, \boldsymbol{a})$. Correspondingly, the transition probability $T$ becomes $T(\boldsymbol{s}, \boldsymbol{s}^{\prime}, \boldsymbol{a})$, where $\boldsymbol{s}^{\prime}$ is the next joint state of all agents. 

In RTB, from a single bidding agent's perspective, each auction involves an unknown number of other bidders, a.k.a opponents. Each bidder has different budget and target preferences. Therefore, in every auction, the set of opponents are highly stochastic. The winner of the auction only observes the highest price from its opponents regardless the bid prices from other bidders. In this way, the stochastic formulation of RTB can be greatly simplified as a two players game. The winning price can be modeled as the joint action from all the other opponents and is partially observable. Unlike other games where the status of the opponents are usually can be seen, in RTB auctions, the opponent's attributes remains unknown.

In the ideal MFE scenario, it supposes that all the agents take a fixed and steady bid distribution $g$ and their own belief of the bid valuation as the prior knowledge to optimize their strategy\cite{iyer2014mean}. The policy that each agent followed is stationary. 
In practice, the bid valuation is estimated by the CTR prediction model and the opponent bid distribution is the market price model. Thus, in this work, we adopted two pre-trained models into the framework to fulfill the above assumption.

We consider the bidding process as an episodic task and each episode consists of $K$ auctions. Each episode has a fixed budget $B =$  CPM\textsubscript{train} $\times 10^{-3} \times K \times c_{0} $, where CPM\textsubscript{train} is the cost per mille impressions in the training data and $c_{0}$ is the budget constraint ratio. 

\begin{figure}[t!]
\centering
\includegraphics[scale=0.35]{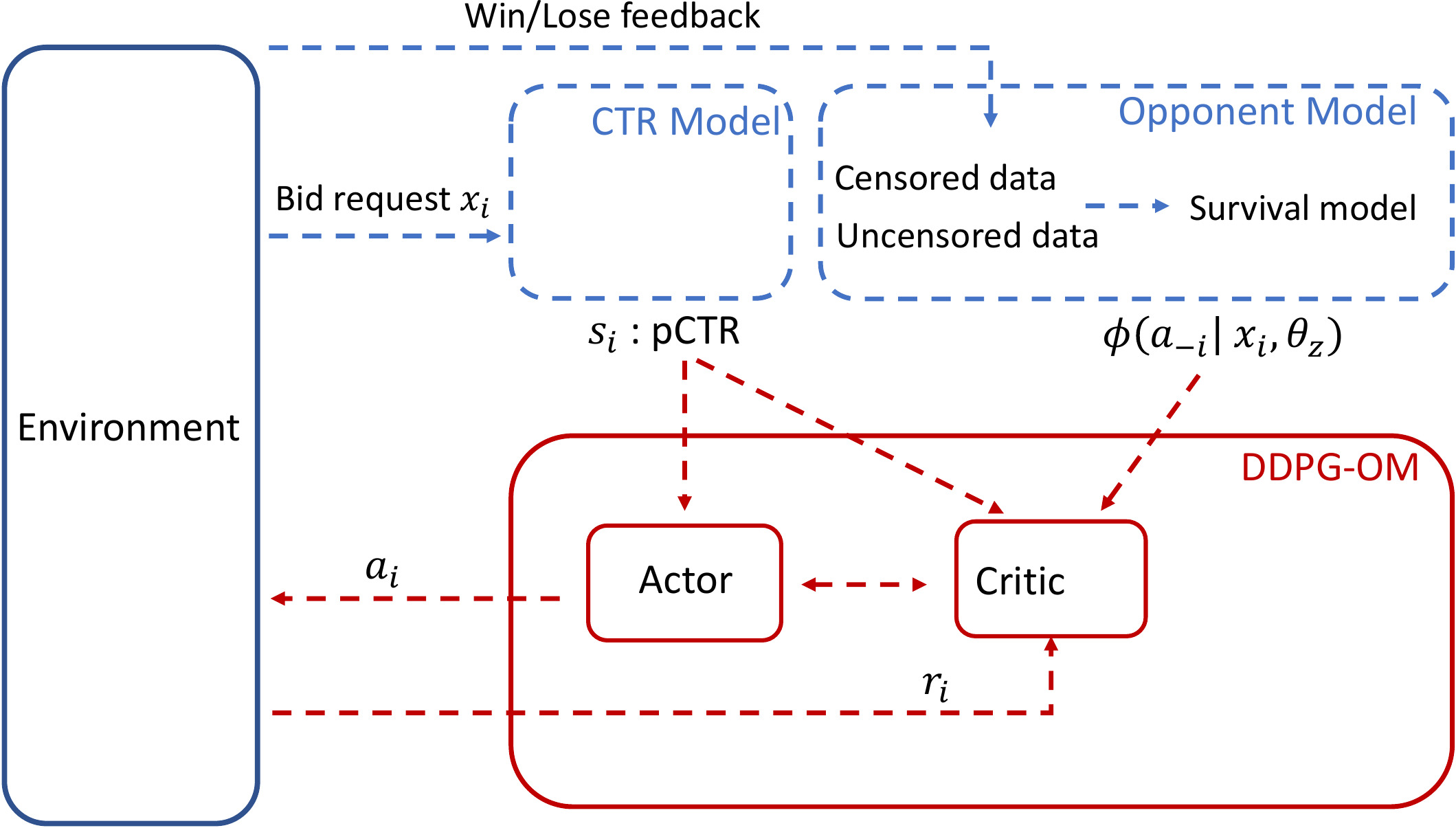}
\caption{The architecture of the DDPG-OM model. The CTR and opponent models are trained offline while the DDPG-OM model is trained online.}
\label{fig:DDPG-OM}
\end{figure}

Fig.~\ref{fig:DDPG-OM} depicts the architecture of the components used in this work. The CTR model takes the feature vector $\textbf{x}$ in the historical bid requests as input and binary labels 0 and 1 indicating impression and click respectively. The predicted Click Through Rate (pCTR) is later used to construct the agent state $s$ and the reward $r$ in the DDPG with Opponent Modeling (DDPG-OM) model. In the following sections, the opponent model and the DDPG-OM model are described in details.

\subsection{Opponent Modeling}
\label{sec:AOM}
In this section, we focus on modeling the opponents actions, a.k.a the market price distribution. The opponent model is defined as the market price distribution at an impression level. We use $a_{-i}$ to represent the action taken by the opponents, a.k.a the highest price from all the other participants in the auction. In this study, $a_{-i}=z$, where $z$ is the market price. The probability density function (P.D.F) of $z$ is $p_{z}(z)$.

\begin{equation}\label{eq:pr}
    p_{z}(z) = P(a_{-i}=z | x, \theta) = h_{z} \prod_{j<z}(1-h_{j})
\end{equation}

As is shown in Eq. \ref{eq:pr}, the P.D.F of the market price can be calculated from the instant hazard function $h_{j}$ which indicates the probability of the instant occurrence of the event at time $j$ conditioned on the event has not happened prior to time $j$. In the RTB setting, $\prod_{j<z}(1-h_{j})$ represents the losing probability of bidding less than the market price and $h_{z}$ shows the probability of observing the market price $z$.

We take the features in the bid request as input and predict the hazard function $h$ over the discretized bid price space at each impression level. The $p_{z}$ can be easily derived from Eq.~\ref{eq:pr}. For the uncensored data, the true label is an one-hot encoded vector of size $b_{max}$ with the element indexed by the market price as 1. 

We followed the loss functions in \cite{ren2019deep}, for the uncensored data, the loss of the observed market price is defined as:
\begin{align}\label{eq:Lz}
L_{z} = -\sum_{x, a_{-i} \in \mathcal{D}_{\text{uncensored}}}[log h_{z} + \sum_{j<z}log(1-h_{j})] \nonumber
\end{align}

For the censored data, it is certain to still lose the auction by bidding lower than the current price. The corresponding loss is defined as:
$$
\begin{aligned}
L_{\text{censored}}
      = -\sum_{x, a_{i} \in \mathcal{D}_{\text{censored}}}\sum_{j<a_{i}} log(1-h_{j}) \nonumber
\end{aligned}
$$

In addition, for the winning auctions, by bidding at any price higher than the observed market price, it is guaranteed to win the auction.
Such information can be shared with the censored data. The loss function is defined as followed:
$$
\begin{aligned}
L_{\text{uncensored}} 
      = -\sum_{x, a_{i} \in \mathcal{D}_{\text{uncensored}}}log[1 - \prod_{j<a_{i}}(1-h_{j})] \nonumber
\end{aligned}
$$

The total loss of the model takes the combination of the above losses as below where $\alpha$ balances the loss values. 
$$
\begin{aligned}
L_{\text{total}}= \alpha L_{z} + (1 - \alpha)(L_{\text{censored}} + L_{\text{uncensored}})
\end{aligned}
$$

%
%
%

\subsection{Bidding Model}
Under our repeated second-price auctions setting, in every auction, all the agents are facing the same bid request. The agents bid for the same ad campaign upon different requests with unknown number of opponents at each auction. The RL agent adopts the framework of Deep Deterministic Policy Gradient (DDPG) \cite{lillicrap2015continuous} method to learn the policy in a continuous space. 

\textbf{State.}
For the DDPG agent, we take the budget left in an episode $B_{i}$ and the $pCTR$ as the state $s = <B_{i}, pCTR>$

\textbf{Action.}
Following the settings in \cite{jin2018real}, the action $a_{i}$ is set to be a scaler which controls the bid price and is bounded to be in the range of $[0, 1]$. The final bid price is calculated by $b_{f} = \text{min}(b_{max} \times a_{i}, B_{i})$, where $b_{max}$ is the upper bound of the bid price. The market price or the aggregated actions from the opponents are denoted as $a_{-i}$.

\textbf{Reward.}
The reward is usually the Key Performance Indicator (KPI) defined by the advertisers, for instance, a click, a purchase or the profits. But such reward signal is usually too sparse for the agent to learn. Therefore, in this study, we assign the $pCTR$ as the reward for all the winning auctions, even without the real click. For the losing auctions, since no price is paid, the reward remains as zero. 
%
%
%
For agent $i$, the actor network takes state $s$, which consists of the predicted CTR and the budget left in the current episode, and parameterized with $\theta_{\pi}$ for a deep neural network which provides an action to take in the range of $[0, 1]$.

\textbf{Action function}\label{eq:actor}
\begin{equation}
a_{i} = \pi_{i}(s_{i}, \theta_\pi) = \pi_{i}([b_{i}, pCTR_{i}], \theta_\pi)
\end{equation}

In the vanilla version of DDPG algorithm, the critic function $Q(s_{i}, a_{i})$ takes the state and action pair from a single agent. In our model, the Q-function is approximated by the mean field theory by integrating the opponent's action distribution. As is shown in Eq.~\ref{eq:critic}, $\phi(a_{-i}|x, \theta_z)$ is the market distribution obtained from the opponent model. The action $a_{-i}$ is not directly observed from the environment, since the result of the auction can only be see after placing a bid price. The market distribution provides the agent's belief of the opponents actions. The indicator function allows the agent to account for the Q value only in the case of bidding higher than the  
market price. Since when the action $a_{i}$ is lower than $a_{-i}$, the agent cannot win such auctions, thus, the Q value should be zero. 

\textbf{Critic function}

\begin{equation}\label{eq:critic}
 \begin{array}{cc}
Q(s, \boldsymbol{a}) = \int_{a_{-i}} Q(s, a_{i}, a_{-i})\phi(a_{-i}|x, \theta_z) \mathds{1}[a_{-i}<a_{i}]\ da_{-i} 
  \end{array}  
\end{equation}



The pseudo code of the DDPG with Opponent-Model (DDPG-OM) algorithm is shown in Alg.~\ref{alg:DDPG-OM}.

\begin{algorithm}[tbh]
\caption{DDPG-OM}\label{alg:DDPG-OM}
Initialize actor network $\pi(s,\theta_\pi)= a_{i}$  and critic network $Q(s,\textbf{a}|\omega)$ with weights $\theta_\pi$, $\omega$ \\
Initialize target network $\pi'$ and $Q'$ with $\theta_\pi' \leftarrow \theta_\pi$ and $\omega' \leftarrow \omega$ \\
Initialize replay memory with size K\;
  \For{episode = 1 to E}{
      receive state $s_{0}$ and sample $a_{0} \sim \pi(s_{0}, \theta_\pi)$\;
      Initialize a noise generator $\mathcal{N}$ for action exploration \\
      \While {$s_{i}$\ not\ terminate}{
        Select an action $a_{i} = \pi(s_{i}, \theta_\pi) + \mathcal{N}_{i}$ and execute \;
        Observe $r_{i}$, $s_{i+1}$\; 
        Store ($s_{i}$, $a_{i}$, $r_{i}$, $s_{i+1}$) in the replay memory\;
        \uIf{$t \equiv 0$ mod $K$}{
    
        sample a minibatch $\mathcal{M}$ from the replay memory \\
        $y_{j} = r_{j} + \gamma Q'(s_{j+1}, \bm{a_{j+1}'}, |\omega') $ \\
        
        update critic by minimizing the loss 
        $L = \frac{1}{\mathcal{M}} \sum_{j}(y_{j} - Q(s_{j}, \bm{a_{j}}|\omega))^2$ \;
        update actor 
        $\theta \leftarrow \theta + \frac{1}{\mathcal{M}} \sum_{j} $\\
        $\nabla_{a} Q(s_{j},\bm{a_{j}}|\omega)|_{s=s_{j}, a=\pi(s_{j})} \nabla_{\theta_{\pi}}\pi(s_{j}|\theta_{\pi})|_{s_{j}}$\;

        update target network: \\
        $\theta'\leftarrow \tau \theta_\pi + (1-\tau)\theta_\pi$\;
        $\omega'\leftarrow \tau \omega + (1-\tau)\omega$
        }
      }
     }
\end{algorithm}

\subsection{Single Agent Steady Market Distribution}
We start from the simplest scenario: a single agent bids against the steady market price distribution. In this setting, we assume the linear bidders have fixed strategies which means they do not update their strategies upon other bidders' actions. In addition, given the dynamic attributes of the bidders, from agent $i$'s point of view, the bids from its opponents are identically and independently distributed. As we discussed in ~\ref{sec:intro}, in practice, the opponent sets in every auction changes over time. Here we assume the departure and the arrival rate of bidders remains steady, which guarantees the stationary of the bid price distribution of the opponents. Even that the opponent bids are partially observable, this allows us to approximate a fixed opponent model and use it in the mean field model. 


\subsection{Multi-Agent Mean Field Approximation}
From the above single agent scenario, here we extend to discuss the mean field equilibrium. 
Instead of considering only one agent, we focus on the multi agent environment, where all agents assume to share one steady bid distribution $\phi$. Taken $\phi$ as the prior knowledge, each agent optimizes their bidding strategy which in turn induces dynamics in the overall bid distribution. 
The mean field equilibrium requires a consistency check of the bid distribution \cite{iyer2014mean}. 
Let $\phi $ be a bid distribution and $\pi_i$ denote a stationary policy for an agent facing bidding decision. 
The mean field equilibrium is achieved if it satisfies the following definition:


\begin{definition}
\label{mfe}
The repeated second-price auction Mean Field games admit at least one \textit{mean field equilibrium} (MFE)\cite{iyer2014mean}, with strategy $\pi$, if:
\begin{enumerate}
  \item $\pi( \cdot | \phi)$ is an optimal strategy given $\phi$.
  \item $\phi$ is the steady state bid distribution given $\pi$. 
\end{enumerate}
\end{definition}

In this game, we assume that the number of competing agents is large.
For each auction, a finite number of agents is randomly selected (through Gaussian noise randomly selected a competing agent, the agent with the largest noise added one is effectively selected). 
Each agent has a random life-time, which is exponentially distributed with unit mean. It optimizes the utility over its lifetime. The unit mean is effectively the fact that each agent starts with the same budget, however the varying lifetime depends on the pCTR values estimated and also the exponentially distributed additive noise. 
At either the end of the episode or when the budget has been exhausted, agents are replaced by new ones whose initial budget, valuation distribution and income is sampled. In most experiments, instead of randomly sampling budget we initialized this to the same value, and noticed no difference in convergence guarantees.
Due to a learning rate decay, eventually the DDPG agent will converge to a stationary agent (learning rate $\approx 0$), thus the normal theorem by \cite{iyer2014mean} holds. 

In the MFE, each bidder are facing i.i.d highest opponent bids and has no incentive to change her bidding strategy. 
However, it is important to note that before the equilibrium is reached, the bid distribution would change as the market evolves. Thus it is important for the agents to infer the bid distribution over time. 

\section{Experiments}
In this work, the experiments are conducted over the public real-world dataset, iPinYou, one of the leading ad companies in China. The dataset contains the original bid logs and the labels of click and purchase. We follow the data pre-processing and feature engineering procedure in \cite{DBLP:journals/corr/ZhangYW14}. Since in the iPinYou dataset, it records the original market price of the impressions, we initiate all the agents with the budget to be proportional to the total cost in the training data: $c_{0}$= 1/16, 1/8, and 1/4. In this way, it allows us to simulate the auctions offline. Given each bid request, each agent in the environment places a bid price and follows the second-price auction principles \footnote{The experiment code will be available for the final version}. The original market price in the log is not included in the environment. 

In Fig~\ref{fig:DDPG-OM}, the CTR estimator is trained offline by adopting the widely used FTRL-logistic regression model \cite{mcmahan2013ad}. In both single and multi agent scenarios, we begin with running the bidding simulation over the training set and log the bid price of each agent and select the second highest price as the market price. The opponent model in Figure~\ref{fig:DDPG-OM} takes the simulated bid log and the features in the original bid requests as input to predict the impression level market distribution as described in Section ~\ref{sec:AOM}. 

Once the CTR model and the opponent model are trained, we repeat the bidding simulation on both train and test set. In this round, the DDPG agent learns the policy while having the prediction of the distribution of its opponent. We begin with setting one DDPG agent in the environment and keep the other bidders using simple and static bidding strategies, for instance, linear bidding function. In this setting, we demonstrate the advantage of the learning agent over the static agent without the learning process. Furthermore, we extend to the multi-agent scenario where all the agents learn their strategies with its estimated opponent model. 

In this study, every 1000 auctions is defined as one epoch. The budget resets at the beginning of each epoch.

\subsection{Opponent Model}\label{sec:OP}
In this section, we compare the general behaviour of the DASA model with other survival analysis models. The experiments are conducted on 3 datasets: Clinic\cite{knaus1995support}, Music\cite{Jing:2017:NSR:3018661.3018719}, and Bidding dataset. The statistics of the datasets can be found in~\cite{ren2019deep}. The data is processed and the results with * in Table~\ref{tab:results} are reproduced in the same way by using their publicly available code \footnote{\url{https://github.com/rk2900/drsa}.} and datasets \footnote{\url{https://www.dropbox.com/s/q5x1q0rnqs7otqn/drsa-data.zip?dl=0}.} and served as baselines in this study. The evaluation metric is the average negative log probability (ANLP) of the market price, which corresponds to the true market price likelihood loss. The result shows the DASA model significantly outperforms other methods across all three datasets.
Thus it is selected as our opponent model for the experiments in the following sections. 

\begin{table}[t]
	\centering
	\caption{Performance comparison on and ANLP}
	\resizebox{\columnwidth}{!}{
		\begin{tabular}{|c||c|c|c|}
			\hline
			\multirow{2}{*}{\textbf{Models}} &  \multicolumn{3}{c|}{\textbf{ANLP}}\cr & CLINIC & MUSIC & BIDDING \cr
			\hline
			KM* & 9.012 & 7.270 & 14.012 \cr
			Lasso-Cox* & 5.307 & 28.983 & 34.941 \cr
			\hline\hline
			Gamma* & 4.610 & 6.326 & 5.941 \cr
			STM* & 3.780 & 5.707 & 4.977 \cr
			MTLSA* & 17.759 & 25.121 & 9.979 \cr
			\hline\hline
			DeepSurv* & 5.345 & 29.002 & 35.405 \cr
			DeepHit* & 5.027 & 5.523 & 5.513 \cr
			DRSA* & 3.337 & 5.132 & 4.598 \cr
			DASA (One Stack Transformer) & \textbf{2.786} & \textbf{4.912} & \textbf{3.465} \cr 
			\hline
		\end{tabular}
		\label{tab:results}
	}
\end{table}



\subsection{Single DDPG agent with Steady Market Distribution}
In this section, we assume that there is one learning agent running DDPG algorithm and competing against $N$ bidders with fixed strategies, for example, a linear bidding function.
In practice, $N$ is always unknown and for each auction, a random set of the $N$ bidders is selected. 
In addition, the set of bidders may at different stage of their lifetime with different budget left. 
In the second-price auction, the most important opponent is the bidder with the second highest price among all the bidders. 
In this study, we set up two linear bidders and one DDPG bidder. The bidders take the same pCTR from the CTR prediction model. 
By injecting Gaussian noises into the pCTR, we simulate the stochastic environment of random bidders with different at each auction. 
The bid price of the DDPG agent are compared with the price generated by the linear bidders and the new market price is logged and used as the input for training the survival model. 
The survival model is trained offline which takes the features $x$ in the bid log and the winning or losing signal from the environment as described in Sec.\ref{sec:OP}. 

\begin{figure}[thb]
\centering
\includegraphics[width=0.35\textwidth]{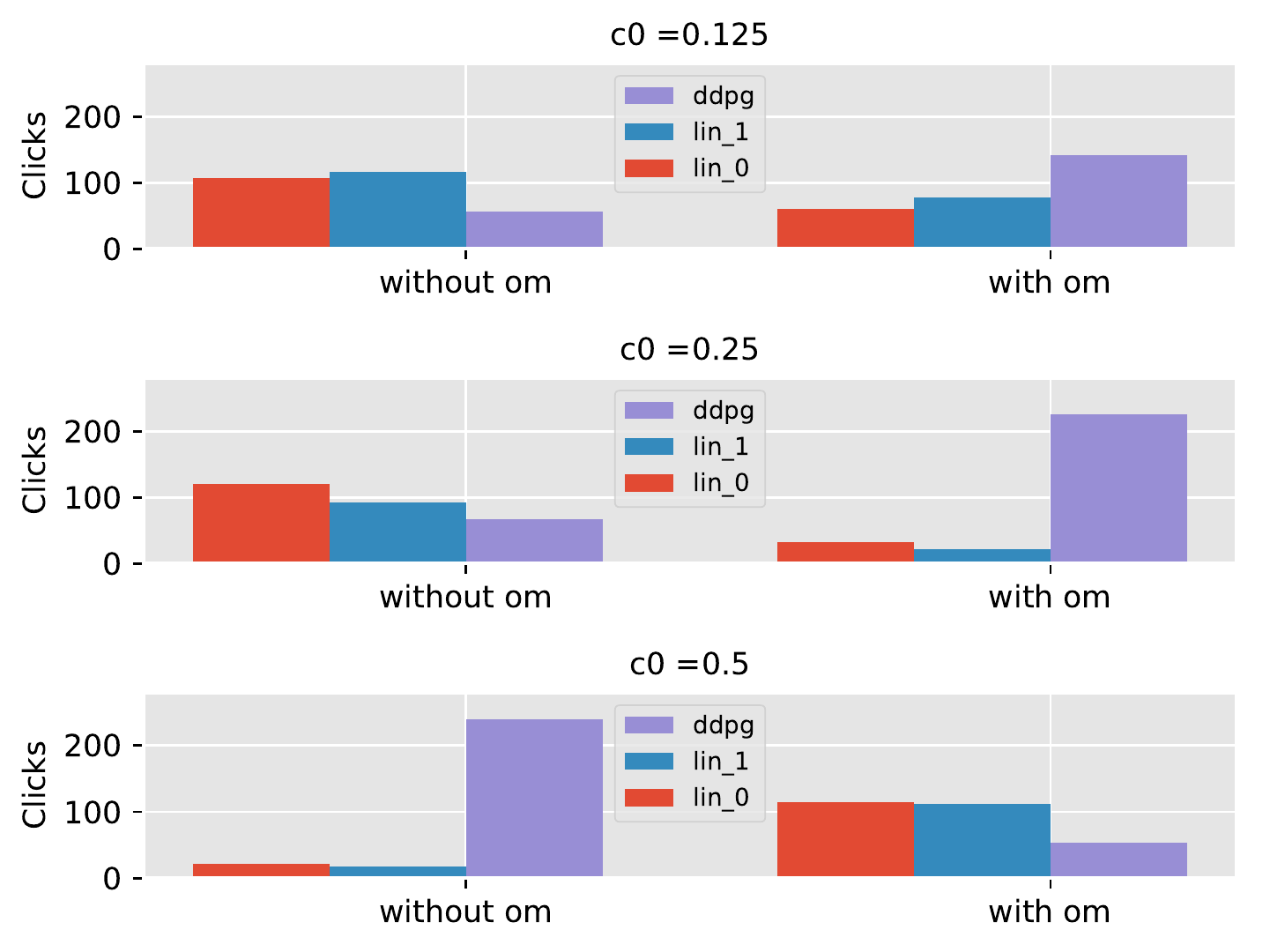}
\caption{camp.2259 winning clicks}\label{fig:2259-clicks}
\end{figure}

%
%
In the next round, we replay the bidding game again to train the same DDPG agent from scratch with the opponent model integrated. In Figure ~\ref{fig:2259-clicks}, the number of clicks obtained by the three bidders are listed for one selected ad campaign, 2259.  

\begin{figure}[thb]
\centering
\includegraphics[width=0.35\textwidth]{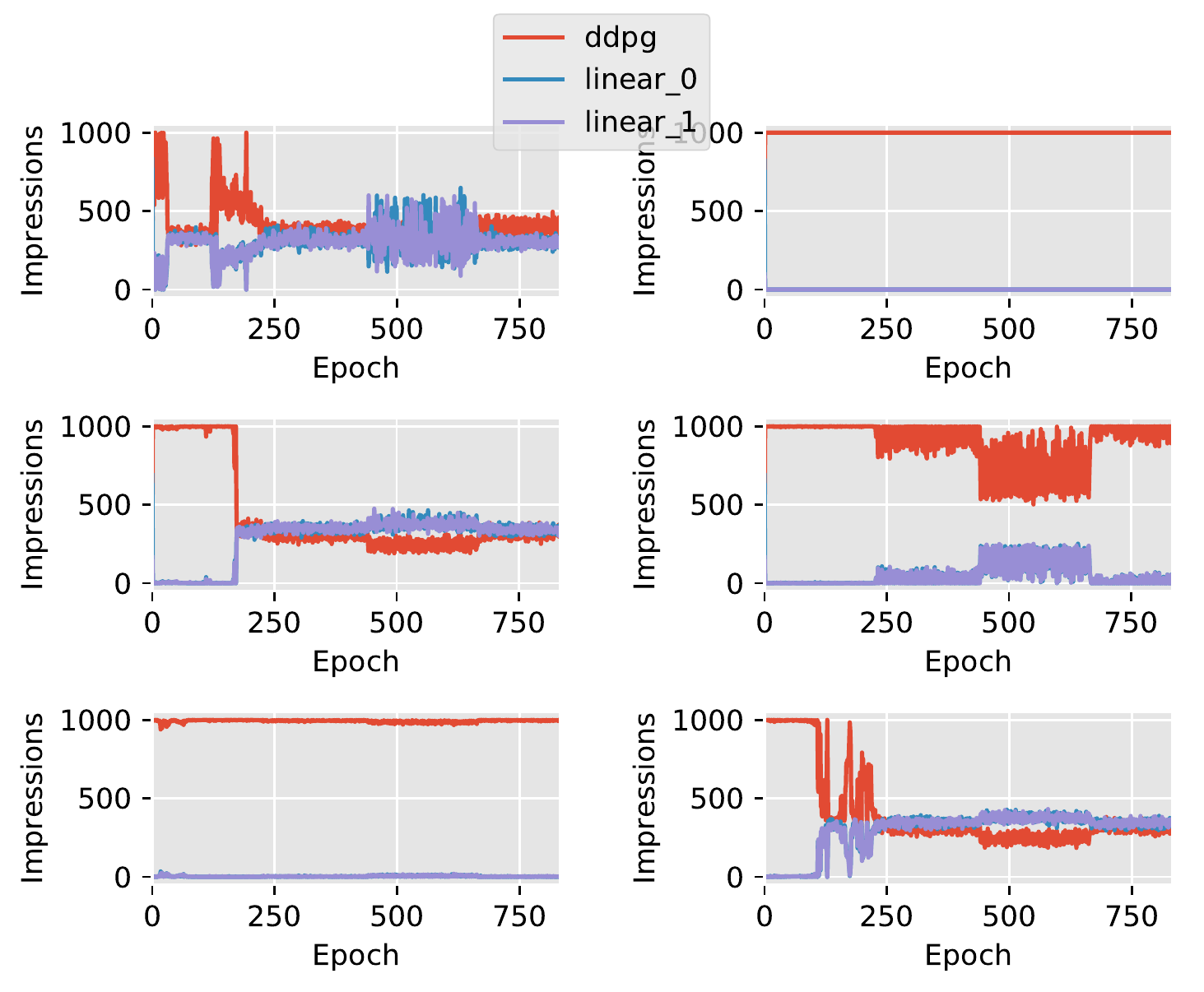}
\caption{camp.2259 winning impressions}\label{fig:2259-imps}
\end{figure}



In Figure ~\ref{fig:2259-imps}, it shows the number of impressions each agent won in each epoch. The rows represents three budget settings, where $c_{0}$= 1/8, 1/4, and 1/2. The left column are the results from DDPG agent without opponent model as the baseline while the right column shows the DDPG agent with opponent model. By having the opponent model, the DDPG agent starts dominating the bidding game. Since the other agents are unaware of the market change, they always bid proportionally to the predicted CTR. 

We need to note that, the budget was set by referring to the original market price in the iPinyou bidding log. But the new market price generated by the agents are different and lower. Thus, it is the reason that for some campaign like 3358, and 2821, even with $c0=0.25$, it is sufficient for the DDPG agent to dominate the other linear bidders without the opponent model. When the DDPG dominates the game, the market price model is approximately fully observable, thus, the gain becomes insignificant. 
At the beginning of their campaign lifetime, without any information of the market, the learning agent converges to a steady but sub-optimal strategy. However, if they infer the opponents model quickly and the opponents have fixed strategy, the DDPG-OM model facilities the bidder to converge to a more dominant strategy in the market. 
If the other agents adopt learning process into their strategies which evolves the bid distribution, the challenge would be to show the asynchronous best response from all the agents and converge to the MFE which is shown in the next section.

\begin{table}[!th]\label{tb:clicks}
\centering
\caption{Clicks gain of DDPG-OM under different budget settings}
\begin{adjustbox}{max width=0.35\textwidth}
\begin{tabular}{|l|l|l|}
\hline
camp. & c0=0.125 & c0=0.25 \\ \hline
2259 & 153\% & 266\% \\ \hline
2821 & 164\% & 2.8\% \\ \hline
3358 & 107\% & -5.4\% \\ \hline
\end{tabular}
\end{adjustbox}
\end{table}

\subsection{Multi-agent game}
\begin{figure}[tbh!]
\centering
\includegraphics[width=0.4\textwidth]{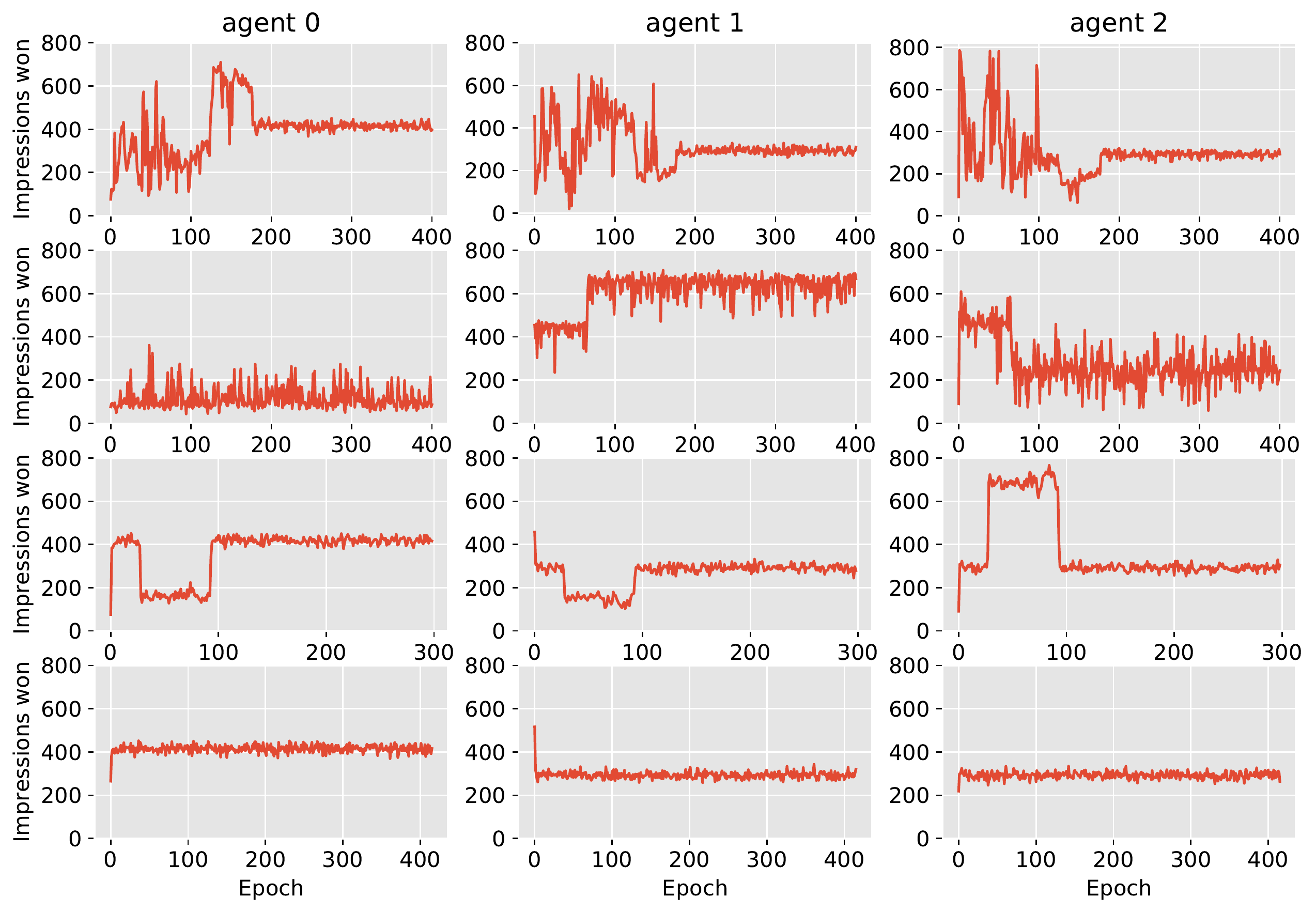}
\caption{3 ddpg agents bidding game. Row 1: without market model. Row 2: with random market model.  Row 3: with DASA model Row 4: test set}\label{fig:2259-marl-win}
\end{figure}

In this section, the experiment is extended to have multiple learning agents in the same environment. As is shown in the first row in Figure~\ref{fig:2259-marl-win}, the 3 agents start with bidding by only learning from its own reward without referring to other bidders' behaviour. After 200 epochs, the game converge to the equilibrium where the number of impressions won by each agent roughly evenly distributed. In the second row, the market distribution per impression is sampled from an uniform distribution. In this case, it increased the variance of the number of the impressions won in each episode and some agent may converge to a dominating strategy. We take the bidding log generated by the first game and trained a market model separately for each agent based on the set of impressions they won. With the information of the market, we reset the game for the training set, as is shown in the third row in Figure~\ref{fig:2259-marl-win}. The agents converge the optimal strategies within 100 epochs which is 50\% less than the results in the first row. We further test the model on the test set, which shows the model is generalized well and the equilibrium is reached.

\section{Conclusions}
In this paper, we propose a general opponent aware bidding algorithm with no prior assumptions on the opponents bidding distribution. To the best of our knowledge, it is the first experimental implementation in the real-time bidding domain to infer the partially observable opponents in the policy learning process. We proposed a deep attentive survival model as the impression level opponent model. The multi-agent bidding simulations show the benefits of improved convergence rates for the DDPG model across all budgets with augmented with an opponent model. For the future work, we will investigate the online training for the opponent model in the multi-agent bidding game.

\section*{Acknowledgments}
This research has been supported by the National Research Fund (FNR) of Luxembourg under the AFR PPP scheme.
The experiments presented in this paper were carried out
using the HPC facilities of the University of Luxembourg~\cite{VBCG_HPCS14}
{\small -- see \url{https://hpc.uni.lu}}, and the facilities provided by MediaGamma Ltd.

\bibliographystyle{named}
{\small \bibliography{ijcai19}}

\end{document}